\def \cm{~\rm{cm}}
\def \s{~\rm{s}}
\def \km{~\rm{km}}

\def \yr{~\rm{yr}}

\documentclass[12pt,preprint]{aastex}
\usepackage{natbib}

\shorttitle{Pairs of bubbles}
\shortauthors{Soker}
\slugcomment{Draft version of \today}

\begin{document}

\title{PAIRS OF BUBBLES IN PLANETARY NEBULAE AND CLUSTERS OF GALAXIES}

\author{Noam Soker\altaffilmark{1}}

\altaffiltext{1}{ Department of Physics, Oranim, and
Department of Physics, Technion-Israel Institute of
Technology, 32000 Haifa, Israel;
soker@physics.technion.ac.il.}

\begin{abstract}

I point to an interesting similarity in the morphology and
some non-dimensional quantities between pairs of X-ray-deficient
bubbles in clusters of galaxies and pairs of optical-deficient
bubbles in planetary nebulae (PNs).
This similarity leads me to postulate a similar formation mechanism.
This postulate is used to strengthen models for PN shaping by jets
(or collimated fast winds: CFW).
The presence of dense material in the equatorial plane observed
in the two classes of bubbles constrains the jets and CFW activity
in PNs to occur while the AGB star still blows its dense wind,
or very shortly after.
I argue that only a stellar companion can account for such jets and CFW.
\end{abstract}


\section{INTRODUCTION} \label{sec:intro}

{\it Chandra} X-ray observations of clusters of galaxies reveal the
presence of X-ray-deficient bubbles in the inner regions of many
clusters, e.g., Hydra A (McNamara et al.\ 2000),
Abell~2052, (Blanton et al.\ 2001, 2003),
A 2597 (McNamara et al.\ 2001), RBS797 (Schindler et al.\ 2001),
Abell~496 (Dupke \& White 2001), and Abell~4059 (Heinz et al.\ 2002). 
These bubbles are characterized by low X-ray emissivity implying low
density (high quality images of some clusters, e.g., A 2597,
Perseus A, and Hydra A, are on the {\it Chandra} home page:
http://chandra.harvard.edu/photo/category/galaxyclusters.html).
In most cases, the bubbles are sites of strong radio emission.
In some cases a pair of radio jets connects the bubble with
the active galactic nucleus (e.g., Hydra A: McNamara et al.\ 2000).
The absence of evidence of shocks suggests that the bubbles are 
expanding and moving at subsonic or mildly transonic velocities
(Fabian et al.\ 2000; McNamara et al.\ 2000; Blanton et al.\ 2001).
The presence of bubbles which do not coincide with strong radio emission
(known as `ghost bubbles' or `ghost cavities')
located farther from the centers of the clusters in, e.g.,
Perseus (Fabian et al.\ 2000), and Abell~2597 (McNamara et al.\ 2001),
suggests that the bubbles rise buoyantly.

The optical morphologies of some planetary nebulae (PNs) reveal 
pairs of bubbles (cavities), similar in morphology to the pairs
of X-ray-deficient bubbles in clusters of galaxies.
Examples are the Owl nebula (NGC 3587; PN G148.4+57.0: 
Guerrero et al.\ 2003), and Cn 3-1 (VV 171; PN G038.2+12.0;
Sahai 2000), which display a pair of low emissivity bubbles along
their symmetry axis. 
{{{ Note that in PNs the bubbles are optical-deficient, and not
X-ray deficient. In PNs the optical deficient bubbles may actually
be filled with X-ray emission, e.g., as in NGC 6543
(Chu et al.  2001). }}}
An interesting object is Hu 2-1 (PN G051.4+09.6), which possesses
two prominent pairs of bubbles, one pair closer to the center and the
other farther out, with inclination between the two symmetry axes
(Miranda et al.\ 2001;
the HST images of Cn 3-1 and Hu 2-1 can be found in the
list compiled by Terzian \& Hajian 2000:
http://ad.usno.navy.mil/pne/gallery.html).
This morphological structure is most similar to the two
prominent pairs of X-ray-deficient bubbles in the Perseus cluster
(Abell~426: Fabian et al.\ 2000, 2002). 
Many other PNs also possess pairs of low emissivity bubbles,
although in most cases they are less prominent than the
cases quoted above, e.g., NGC 2242 (PN G170.3+15.8:
Manchado et al.\ 1996), and a pair of bubbles within bipolar
lobes in M 2-46 (PN G024.8-02.7: Manchado et al.\ 1996). 
{{{ Some similar bubbles images of PNs and clusters are listed 
in Table 1. }}}

The morphological similarity between X-ray-deficient bubbles
in clusters of galaxies and optical-deficient bubbles in
PNs hints at a similar formation process in these vastly
different objects.
In Section 2 I explore the common and different properties
of bubbles in these two classes of objects.
The goal is to learn about the formation process of
bubbles in PNs from information available from clusters
of galaxies.
In particular, I argue in Section 3 that the commonly accepted
model for the formation of bubble pairs in clusters of galaxies,
where the energy is injected in two oppositely propagating jets,
strongly hints that the same mechanism operates in PNs.
This similarity further constrains the object blowing the jets
in PNs to be a compact object, and to blow the jets during the
asymptotic giant branch (AGB) phase of the progenitor or during
the early post-AGB phase.
I argue that the compact object must therefore be a companion.
A short summary is in Section 4.

\section{COMPARISON OF BUBBLE PROPERTIES} 
\label{proper}

Some properties of bubbles in clusters of galaxies and in PNs are
compared in Table 2.
The properties of the bubbles are taken from the observational
papers cited in the previous section and the theoretical studies
cited in the next section.
We note that although the general structure in PNs is observed in
the optical, inner cavities, not necessarily in pairs of bubbles,
may be observed in X-ray (e.g., in the PN NGC 7009;
Guerrero, Gruendl, \& Chu 2002).
Those observations are used to estimate the temperature inside
pairs of bubbles used in table 2. 
The several orders of magnitude difference in some quantities
are obvious from table 2.
The main qualitative differences between the two classes are:
(1) The bubbles in clusters evolve inside the intracluster medium
(ICM), which is in hydrostatic equilibrium; if global flow is present,
it is highly subsonic.
The bubbles in cluster moves outward because of buoyancy. 
In PNs the bubbles move outward as part of the global outflow
of the wind. Gravity is negligible in PNs.
(2) In clusters the inflated bubble $PdV$ work goes mainly to
push material against the high pressure surroundings,
while in PNs the $PdV$ work mainly goes to accelerate AGB wind
material to higher velocities.

The relevant similarities between the two types of bubble pairs are
as follows.
(1) The most relevant similarity is in the morphological structure
(see previous section and table 1 for references to high quality
images).
In particular, in many cases there is a dense region in
the equatorial plane between the two bubbles,
e.g., the cluster A 2597 (McNamara et al.\ 2001)
and the Owl PN(NGC 3587: Guerrero et al.\ 2003).
(2) In some cases more than one pair of bubbles are seen,
e.g., in the Perseus cluster (Fabian et al.\ 2000, 2002) and in
the PN Hu 2-1 (Miranda et al.\ 2001).
(3) In both types of bubbles the density inside the bubble is
2-3 orders of magnitude lower than that in the environment,
with an opposite ratio in temperatures.
(4) In both cases the typical lifetime of observed bubbles is
{{{ estimated to be }}} $\sim 10$ times the estimated duration 
of the main energy injection phase that forms the bubbles.
{{{ This value is highly uncertain, and may vary a lot from 
one system to another. 
In many PNs the linear increase of outward velocity with distance
along the symmetry axis hints that the ejection phase was indeed
of short duration compared with the age.  }}} 
(5) In clusters the bubbles are moving subsonically, or mildly
supersonically, through the ICM.
In PNs the situation is more complicated.
Before ionization by the central star or a companion, the sound
speed in the AGB wind is $\sim 1 \km \s^{-1}$, while the bubbles
expand through the AGB wind at a relative speed of
$\sim 10-30 \km \s^{-1}$, which is highly supersonic.
After ionization, the flow is mildly supersonic or even subsonic.
In any case, even during the supersonic phase, the AGB wind
is shocked to a temperature much below the bubble's temperature,
i.e., the expansion velocity is much below the sound speed
inside the bubble. 
Hence this does not change much the overall characteristic of
the bubble flow.
Namely, in both types of bubbles the flow speed is
much below the sound speed inside the bubble.

\section{POSSIBLE IMPLICATIONS OF MORPHOLOGICAL SIMILARITIES}
\label{bubble}

In this section I postulate that the more or less spherical
(fat) bubble formation mechanism in
clusters of galaxies and in PNs is similar.
In clusters there is more information available on the
formation mechanism than in PNs.
I will use these known and/or well accepted properties
of clusters to project on the formation mechanism in PNs.

It is commonly accepted that pairs of bubbles in clusters are
formed by axisymmetric energy injection by an AGN, where most of
the energy is deposited by two jets at two opposite off-center
locations (e.g., Brighenti \& Mathews 2002;
Br\"uggen 2003; Br\"uggen et al.\ 2002; Fabian et al.\ 2002;
Nulsen et al.\ 2002; Quilis, Bower, \& Balogh 2001;
Soker, Blanton, \& Sarazin 2002;
{{{ Omma et al.\ 2003). }}}
An axisymmetrical density structure of the ambient medium is
not needed to form the cluster's bubbles (only very close to the AGN,
on a scale much smaller than the bubble size, does the accretion
disk influence the flow).
This hints, according to the postulate made here, that the bubbles
in PNs are also formed by jets.
The idea that jets, or collimated fast winds (CFW), shape PNs is not
new.
Jet shaping was proposed by several authors to explain different
morphological features, e.g.,
jets (or CFW) blown by a stellar companion (Morris 1987;
Soker \& Rappaport 2000) to explain bipolar PNs, and jets blown at
the final AGB phase or early post-AGB phase to form dense blobs
along the symmetry axis (Soker 1990; these blobs are termed ansae, or
FLIERs for fast low ionization emission regions), or
shape the PN (Sahai \& Trauger 1998).
Recently, more quantitative analyses of bubble inflation by
jets in PNs were conducted analytically (Soker 2002),
and numerically (Chin-Fei \& Sahai 2003). 
However, unlike in clusters, this idea is controversial, with
other models suggesting magnetic shaping or axisymmetrical AGB winds
{{{ (see the debate of Bujarrabal et al. 2000). }}}
The latter two processes can work as well, in particular in parallel with
jets shaping.
Livio (2000) reviews properties of jets in PNs, and compares them
with those of other systems known to blow jets. 
The postulate made here provides insight into the shaping mechanism.

There are other interesting similarities in the bubble morphology
between some clusters and some PNs.
The density in the jets and the ambient medium decrease with increasing
distance from the center.
If the jet in a PN is shocked, or shocks the ambient medium, i.e.,
the AGB wind, close to the center, the dense post-shock regions
cool fast, and no bubble is formed.
Only when the cooling time of the shocked gas is long a bubble is
formed (Soker 2002).
In PNs, bubles are typically expected to be formed by jets quite close
to the center, at distances of $z \lesssim 10^{16} \cm$
from the center (eqs. 6 and 14 in Soker 2002).
The two opposite bubbles expand to all directions, including the
center, and when the PN emerges, it is not possible to observe the
inner region where the jets expanded without forming a bubble;
either this region was destroyed or it is not resolved.
However, for slow, speeds of $\sim 200 \km \s^{-1}$, and/or dense
jets and winds, the post-shock gas will inflate bubbles only at large
distances from the center, $z \sim 10^{17} \cm$ (eq. 6 in Soker 2002).
The expanding bubbles will not destroy the large inner region, where the
jets expand without forming bubbles.
This inner region will be observed as a dense region with two
low density cones (or cylinders), one in each opposite direction,
along which the two opposite collimated jets expanded. 
According to this flow structure, the bubble will be connected to these
cones (or cylinders) at their ends.
Such a flow structure is clearly observed in the cluster Hydra A
(McNamara et al. \ 2000), where two well collimated radio jets
are abruptly shocked and form radio lobes.
The X-ray intensity is high along the boarders of the radio lobes.
In PNs no such nice jets, as the radio jets in Hydra A, are observed.
However, in some PNs the optical morphology is very similar to the X-ray
morphology of Hydra A, suggesting a similar flow structure.
Again, the PN phase comes long after the jets ceased, unlike in Hydra A
where the radio jet is still active, hence the narrow cones through
which the jets expanded were broadened by the fast wind blown
by the central star, and are difficult, or impossible, to identify .
In any case, possible PNs in which the jets inflated bubbles
at large distances from the center are M 1-59 (PN G023.9-02.3,
image in Manchado et al.\ 1996), and NGC 7026 (PN G 089.0+00.3;
H$\alpha$ image in Balick 1987; see also Terzian \& Hajian 2000).
In these two PN the inner region is bright, with a faint narrow cylinder
along the symmetry axis. The two bubbles in each of these PNs 
seem to start from the outer boundary of these inner regions,
rather than from the center.
Yet, another interesting similarity is that the central engine, star(s)
in PNs and an AGN in clusters, may not be exactly on the symmetry axis.
Examples for the departure of the central engine from the symemtry axis
are in the Perseus cluster and the Owl nebula (NGC 3587). 

The similarity in several non-dimensional quantities found in the previous
section suggests that if the initial flow structure is similar,
then the bubble morphologies will be similar, as observed.
This leads to the following.
(1) The similar shapes strengthens the general
idea that jets (or CFW) form and shape the bubbles in PNs, as
well as other types of bipolar PNs. 
(2) The low density in the bubble implies that the jets are fast,
with a speed of $> 100 \km \s^{-1}$.
Therefore, the object launching the jets must be compact,
since the jets speed is of the order of the escape velocity
(Livio 2000).
(3) The presence of more than one pair of bubbles in the PN Hu 2-1
 indicates, as in clusters, multiple episodic events. 
(4) In clusters the surrounding density increases as radius decreases
down to the center.
The similar bubble morphologies and the presence of dense material
in the equatorial plane between the two bubbles (see previous section)
suggests that a similar ambient medium exists in PNs when the jets are
blown.
Namely, the AGB dense wind is still active, or has ceased only
recently, when the jets are blown in PNs.
This is possible only if the jets are blown by a companion, or the
central star moves extremely rapidly from the AGB to become a compact
star that can blow fast jets.
This rapid evolution is in contrast with stellar evolution studies,
and is also unlikely to explain the multiple activity (point 3 above;
Miranda et al.\ 2001 argue for a CFW that was blown by a
binary system progenitor of Hu 2-1).
One of the observational implications is that we should see evidence
for fast jets in objects that are still unambiguous AGB stars.
A good example is the system OH231.8+4.2 (Rotten Egg nebula),
for which Kastner et al.\ (1998) detect the presence of a Mira inside this
bipolar nebula {{{ which contains jets (Zijlstra et al.\ 2001). }}} 
{{{ Zijlstra et al. (2001) present evidence for jets in some OH/IR
early post-AGB stars. 
There are also resolved jets in some AGB stars, e.g., 
W43A (Imai et al.\ 2002, 2003) and V Hydrae (Sahai et al. 2003). }}}

The first three points above, of fast jets, or CFW, shaping PNs,
sometimes in multiple activity, were mentioned before; see the recent
papers by Soker (2002) and Chin-Fei \& Sahai (2003) for
discussions of the arguments for CFW shaping.
The comparison with clusters' bubbles strengthens these points
for PNs having pairs of well defined bubbles.
Point 4 above, is new.
It is unique in strongly hinting at CFW shaping occurring while the
primary still blows its AGB wind.
This implies that the CFW is blown by a companion.

\section{SUMMARY} \label{sec:conclusion}

The purpose of this paper is twofold.
Firstly, to point to a similarity in the morphology and
some non-dimensional quantities between pairs of bubbles in clusters
of galaxies and in PNs (section 2).
In the latter group I considered mainly PNs harboring a pair or more
of well defined and closed bubbles (i.e., fat bubbles).
Examples are given in section 1 {{{ and table 1. }}}
This similarity is interesting by itself, considering the huge
differences in temperatures, size, mass, energy, etc., between the
two groups.
Secondly, I used this similarity to strengthen models
for PN shaping by jets (or collimated fast winds: CFW), and
to constraint the formation epoch of the CFW.

Arguments for shaping of PNs by jets and CFW were presented before
(references in section 3).
{{{ Other effects, though, e.g., enhanced equatorial mass loss rate,
can also play a role in shaping PNs. }}}
It was also assumed that similar mechanisms, e.g., accretion disks,
launch jets in AGN, which shape bubbles in clusters, and in PNs
(Livio 2000).
Here I further postulate similar bubble shaping in cluster of galaxies
and in PNs.
This allows projection from known properties and processes in clusters
to PNs.
My main conclusions based on the similarities are as follows (section 3).
The similarity in morphology and some properties strongly supports
jets or CFW models for the shaping of pair of bubbles in PNs.
The ambient medium in PNs, which is the slow AGB wind, need not be
axisymmetrical, and may be spherical.
The presence of dense material in the equatorial plane constrains
the jets and CFW activity to occur while the AGB star
still blows its dense wind, or very shortly after.
The requirement that the jets and CFW be fast and the presence
of more than one pair of bubbles in, e.g., Hu 2-1,
constrains the object that blows the jets and CFW to be a compact
companion, i.e., a main sequence or a white dwarf star.

Although I considered here only PNs with well defined pairs
of closed bubbles, the results are more general in strengthening
the idea that bipolar and extreme elliptical PNs are shaped by
jets or CFW blown by an accreting companion.

\acknowledgements
{{{ I thank Albert Zijlstra (the referee) 
and Joel Kastner for useful comments, and James Binney for
encouraging comments. }}}
This research was supported in part by the
Israel Science Foundation.

\newpage
\noindent {\bf A comment to table 1:}
Free access to images are at these sites:
\newline
[1] http://arxiv.org/PS\_cache/astro-ph/pdf/0210/0210054.pdf
\newline
[2] http://ad.usno.navy.mil/pne/images/rob22.jpg
\newline
[3] http://arxiv.org/PS\_cache/astro-ph/pdf/0007/0007456.pdf
\newline
[4] http://arxiv.org/PS\_cache/astro-ph/pdf/0303/0303056.pdf 
\newline
[5] http://arxiv.org/PS\_cache/astro-ph/pdf/0107/0107221.pdf
\newline
[6] http://ad.usno.navy.mil/pne/images/vv171.jpg
\newline
[7] http://arxiv.org/PS\_cache/astro-ph/pdf/0010/0010450.pdf
\newline
[8] http://ad.usno.navy.mil/pne/images/he2\_104.jpg
\newline
[9] http://chandra.harvard.edu/photo/cycle1/hcg62/index.html
\newline
[10a] http://arxiv.org/PS\_cache/astro-ph/pdf/0009/0009396.pdf
\newline
also: [10b] http://ad.usno.navy.mil/pne/images/hu21\_ha.gif 
\newline
[11] http://arxiv.org/PS\_cache/astro-ph/pdf/0001/0001402.pdf 
\newline
[12] http://ad.usno.navy.mil/pne/images/ngc6537.jpg
\newline
[13] http://arxiv.org/PS\_cache/astro-ph/ps/0109/0109488.f1.gif
\newline
[14a] http://ad.usno.navy.mil/pne/images/ngc7009.jpg
\newline
see also (Goncalves et al.\ 2003, fig. 1)
\newline
[14b] http://arxiv.org/PS\_cache/astro-ph/pdf/0307/0307265.pdf

\begin{deluxetable}{lll}
\tablewidth{0pt}
\tablecaption{Similar images of PNs and clusters}
\tablehead{
\colhead{Structure} & \colhead{Clusters} & \colhead{PNs}
}

\startdata
Butterfly shape of the & Abell 478        & Roberts 22          \\
bright region; faint   & (Sun et al. 2003, & (Sahai et al. 1999,  \\
along symmetry axis    & fig 1) [1]       & fig. 1a) [2]        \\
\hline
Pairs of fat spherical & Perseus          & NGC 3587   \\
bubbles near center    & (Fabian et al.   & (Guerrero et al. 2003, \\
                       &  2000) [3]       &  fig. 1) [4]          \\
\hline
Closed bubbles         & Abell 2052            &   VV 171         \\
connected at the       & (Blanton et al. 2001,  & (Sahai 2001)   \\
equatorial plane       &  fig. 3) [5]         &   [6]          \\
\hline
Open bubbles           & M 84                  &  He 2-104           \\
connected at the       & (Finoguenov \& Jones  & (Sahai \& Trauger,  \\
equatorial plane       & 2001, fig 1) [7]       &  1998) [8]          \\
\hline
Pair of bubbles     & HCG 62             &  Hu 2-1                \\
detached from a     & (Vrtilek et al.    &  (Miranda et al. 2001,  \\
bright center       &  2002) [9]         &  fig. 2) [10]          \\
\hline
point-symmetric     & Hydra A              & NGC 6537          \\
elongated lobes     & (McNamara et al.     & (Balick 2000,    \\
                    &  2000, fig. 1) [11]  & fig. 2) [12] \\
\hline
Pairs of bright     &  Cygnus A             &  NGC 7009       \\
bullets along the   &  (Smith et al.        &  (Balick et al.         \\
symmetry axis       &   2002, fig. 1) [13]  &   1998, fig. 1,4) [14]  \\
\enddata
\tablecomments{
Similr images of bubbles in clusters of galaxies and 
planetary nebulae (PNs).
In clusters these are X-ray images (e.g., with X-ray deficient bubbles), 
while in PNs they are optical images (e.g., with optical
deficient bubbles). 
In the first five pairs of images the similarity is of high
degree. In the last two pairs of images the similarity
between the cluster and the PN is of lesser degree. 
}
\end{deluxetable}

\begin{deluxetable}{lcc}
\tablewidth{0pt}
\tablecaption{Bubbles and Environment Properties}
\tablehead{
\colhead{Property} & \colhead{Clusters} & \colhead{PNs}
}

\startdata
Environment: type   & ICM              & AGB  wind              \\
Environment: status & hydrostatic      & outflow                \\
Observation         & X-ray            & optical                \\
Size  (cm)          & $\sim 10^{23}$   & $\sim 10^{17}-10^{18}$ \\
$T_{\rm bub}$ (K)      & $> 10^9$   & $\sim 10^6-10^7$       \\
$T_e$  (K)          & $\sim 10^7$      & $\sim 10^4$            \\
$n_e$               & $0.1-0.01$       & $\sim 10^4$            \\
Flow: type          & buoyancy          & outflow              \\
Flow: speed (km$\s^{-1}$) & $\sim 10^3$ & $\sim 10-30$         \\
$\tau_{\rm age}$ (yr) & $\sim 10^7-10^8$ & $\sim 10^3-10^4$       \\
$\tau_{\rm inj}$ (yr) & $\sim 10^7$      & $\sim 100-1000$        \\
$E_b$ (erg)            & $\sim 10^{59}$   & $\sim 10^{45}-10^{46}$         \\
$\dot M_b$ ($M_\odot \yr^{-1}$) &$< 10 $ & $\sim 10^{-5}-10^{-6}$  \\
$T_{\rm bub}/T_e$     &  $\gtrsim 100$   & $\sim 100-1000$        \\
$V_{\rm flow}/c_s$    &  $\sim 1-3$        & $\sim 3-30$             \\
\enddata

\tablecomments{The properties of bubbles and their environment
in clusters of galaxies and planetary nebulae (PNs).
The quantities in the table:
ICM: intracluster medium; AGB: asymptotic giant branch;
$T_{\rm bub}$ and $T_e$: temperatures of the gas inside and outside
the bubble, respectively; 
$n_e$: electron density in the environment of the bubble;
$c_s$: sound speed in the environment;
$\tau_{\rm age}$: the typical age of observed bubbles; 
$\tau_{\rm inj}$: the estimated duration of the energy
injection phase to inflate the bubble;
$\dot M_b$: the estimated mass injection rate into the bubble
during the formation phase.
$E_b$: the energy required to inflate the bubble.
}
\end{deluxetable}


\end{document}